\documentclass[sigplan]{acmart}
\acmSubmissionID{187}
\renewcommand\footnotetextcopyrightpermission[1]{}
\setcopyright{none}
\acmConference[EuroSys '27]{The 22nd European Conference on Computer Systems}{April 19--24, 2027}{Rabat, Morocco}
\acmYear{2027}
\acmISBN{}
\acmDOI{}

\usepackage{booktabs}
\usepackage{enumitem}
\usepackage{tabularx}
\usepackage{tikz}
\usepackage{placeins}
\usepackage{xspace}
\usetikzlibrary{arrows.meta,positioning,fit,shapes.geometric,calc}

\setlist[itemize]{leftmargin=*,topsep=2pt,itemsep=1pt,parsep=0pt,partopsep=0pt}
\setlist[enumerate]{leftmargin=*,topsep=2pt,itemsep=2pt,parsep=0pt,partopsep=0pt}
\newcommand{\system}{Cordon\xspace}
\newcommand{\stx}{semantic transaction\xspace}
\newcommand{\stxs}{semantic transactions\xspace}
\newcommand{\resultobj}{result object\xspace}

\newcommand{\effectobj}{effect object\xspace}

\begin{document}

\title{Cordon: Semantic Transactions for Tool-Using LLM Agents}

% \author{Zheng Chen, Hanqin Liu, Duling Xu, Jialin Li, Dong Dong, and Bangzheng Pu}
% \affiliation{%
%   \institution{AetherHeart Tech Co., Ltd.}
%   \city{Shanghai}
%   \country{China}
% }

\author{Zheng Chen}
\email{zheng-chen@mail.tsinghua.edu.cn}
\affiliation{%
  \institution{Tsinghua University}
  \city{Beijing}
  \country{China}
}

\author{Hanqing Liu}
\email{hanqingliu@sjtu.edu.cn}
\affiliation{%
  \institution{Shanghai Jiao Tong University}
  \city{Shanghai}
  \country{China}
}

\author{Duling Xu}
\email{xuduling@ruc.edu.cn}
\affiliation{%
  \institution{Renmin University of China}
  \city{Beijing}
  \country{China}
}

\author{Dong Dong}
\email{dongd@tsinghua.edu.cn}
\affiliation{%
  \institution{Tsinghua University}
  \city{Beijing}
  \country{China}
}

\author{Jialin Li}
\email{jialin-li@tsinghua.edu.cn}
\affiliation{%
  \institution{Tsinghua University}
  \city{Beijing}
  \country{China}
}

\author{Bangzheng Pu}
\email{bzpu@aetherheart.com}
\affiliation{%
  \institution{AetherHeart Tech Co., Ltd.}
  \city{Shanghai}
  \country{China}
}

\author{Jidong Zhai}
\email{zhaijidong@tsinghua.edu.cn}
\affiliation{%
  \institution{Tsinghua University}
  \city{Beijing}
  \country{China}
}
% \email{{zchen,hqliu,dulingxu,lijialin,dongd,bzpu}@aetherheart.com}

\begin{abstract}
Tool-using LLM agents are shifting the unit of computation from explicit human-issued commands to model-driven tasks with stateful consequences.
Yet today's agent runtimes still expose tools as isolated RPCs.
This interface gives runtimes a convenient integration point, but it lacks a task-scoped execution boundary for commit, rollback, recovery, and audit across multi-step agent workflows.
We argue that this mismatch calls for a runtime containment boundary rather than another per-call guardrail.

This paper introduces \system, a transactional runtime system for staging and validating irreversible agent effects before commit.
A \stx is a task-level execution boundary that binds tool intents and runtime-tracked result lineage to reversible local state, staged external effects, delegated authority, and audit metadata.
\system implements this abstraction with a transaction manager that tracks derived result objects, executes reversible mutations in shadow state, stages outward-facing actions in an effect outbox, and records recovery metadata.
The runtime then validates the composed execution flow before it commits state or releases external effects.
Our evaluation across adversarial and benign workflows shows that \system exposes cross-step violations missed by existing defenses.
It also reduces irreversible-effect failures while preserving benign task completion with modest approval and latency overhead.
\end{abstract}

% \begin{CCSXML}
% <ccs2012>
%    <concept>
%        <concept_id>10011007.10011006.10011050</concept_id>
%        <concept_desc>Software and its engineering~Runtime environments</concept_desc>
%        <concept_significance>500</concept_significance>
%        </concept>
%    <concept>
%        <concept_id>10010520.10010521.10010537</concept_id>
%        <concept_desc>Computer systems organization~Dependable and fault-tolerant systems and networks</concept_desc>
%        <concept_significance>500</concept_significance>
%        </concept>
%    <concept>
%        <concept_id>10002978.10003014.10003017</concept_id>
%        <concept_desc>Security and privacy~Systems security</concept_desc>
%        <concept_significance>300</concept_significance>
%        </concept>
%  </ccs2012>
% \end{CCSXML}



\maketitle

\section{Introduction}

We are at a crossroads in which the primary operator of computing systems is beginning to shift from humans issuing explicit commands to LLM-driven agents acting on their behalf~\cite{yao2023react,schick2023toolformer}.
Equipped with tools, these agents have already delivered substantial productivity gains across software development~\cite{yang2024sweagent}, office and enterprise workflows~\cite{drouin2024workarena}, scientific research~\cite{boiko2023coscientist,bran2024chemcrow}, and engineering design~\cite{liu2026idesigngpt}.
This progress has been driven in large part by the aggressive integration of external tools into agent systems.
At the same time, these agents remain model-driven systems that act under uncertainty.
This creates a new systems challenge around how runtimes mediate irreversible side effects across long-running multi-step tasks.

\begin{figure}
  \centering
  \includegraphics[width=\linewidth]{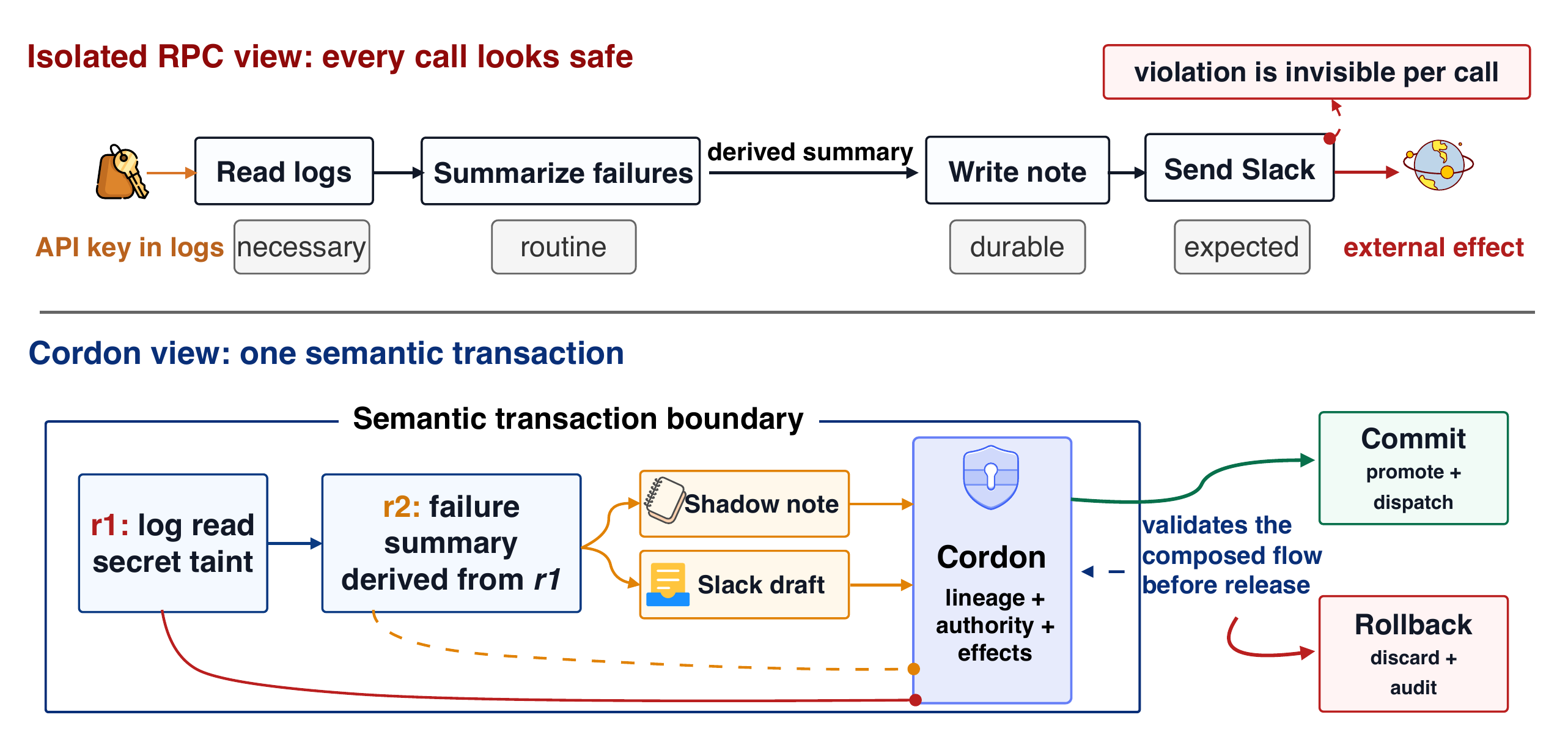}
  \caption{A semantic transaction gives agent runtimes a commit boundary over result lineage, staged mutations, pending effects, and authority.}
  \label{fig:motivating-flow}
\end{figure}

Today's agent runtimes typically expose tools through an RPC-like request/response abstraction.
In this model, the runtime treats each tool request as an independent operation.
It checks or approves the request, executes the tool directly against the underlying system, and loads the result into the agent's memory for subsequent steps.
This interface is convenient for tool integration, but it gives the runtime the wrong execution boundary for long-running agent tasks.
Approving an action or recovering from a failure often depends on the composed execution flow rather than on any single call.
When the runtime records only isolated tool calls, these relationships remain implicit.

Consider an incident-response agent asked to diagnose an outage, as shown in Figure~\ref{fig:motivating-flow}.
It reads application logs that contain an API key, runs commands to summarize failures, writes a remediation note, and prepares a Slack notification for the on-call channel.
Each individual tool call can be justified: reading logs is necessary, shell commands are routine, and notifying the team is expected.
The problematic behavior appears only in the composed task flow, where a secret-bearing result is transformed into a derived summary and then used by an external effect.

We propose \system, a \textbf{\textit{transactional execution runtime}} for tool-using agents.
Rather than validating each tool call in isolation, \system represents each agent task as a \stx and delays irreversible effects until task-level validation becomes possible.
A \stx establishes a \textbf{\textit{commit and recovery boundary}} over result objects, speculative local mutations, pending external effects, delegated authority, and audit metadata.
We use the term operationally to describe a task-scoped runtime commit boundary rather than a formal semantic execution model.
As tools execute, \system materializes result objects and records the lineage through which later operations derive state or effects from earlier results.
This execution model lets \system enforce \textbf{\textit{runtime containment properties}} over flow containment, commit discipline, rollback correctness, boundary mediation, and audit completeness before effects become durable or externally visible.

Operationally, \system interposes at the \textbf{\textit{tool-dispatch boundary}} and executes tool effects transactionally rather than immediately.
A transaction manager translates tool calls into task-scoped intents and attaches each result object to the active transaction context.
Reversible local mutations execute speculatively in \textbf{\textit{shadow state}}, while outward-facing actions are staged in an \textbf{\textit{effect outbox}} and transaction metadata is appended to a recovery log.
At validation time, \system evaluates lineage, authority, staged state, and pending effects as one composed execution flow before releasing external actions.
After validation, \system either commits approved shadow state atomically or aborts staged mutations and effects.
For effects that cross an external boundary after release, \system records the boundary state needed for audit, compensation, and recovery.

We evaluate \system on 45 risk-bearing multi-tool workflows, five deterministic rollback trajectories, and two standard benign agent benchmarks across coding, incident response, document, office, support, and data tasks.
Plain execution commits policy-violating effects in 45/45 risk-bearing workflows; strategy adapters derived from existing defense boundaries intercept only 14/45 before commit and leave 31/45 policy-violating or post-hoc-only, while \system intercepts 45/45 before commit.
With approval wait excluded, transaction-mediated execution reduces mean task time by 24.6--27.9\% relative to plain execution, reduces token use by 23.6--28.4\%, provides 4.17ms median rollback latency with 15/15 resume checks passing, and preserves standard-benchmark task correctness within measurement variance.

We summarize our contributions as follows:
\begin{enumerate}[leftmargin=*]
  \item \textbf{\underline{\emph{New insight.}}} We identify a missing transaction boundary between model-driven tool use and durable side effects. This boundary helps explain why per-call defenses can miss failures that arise only across result lineage, local mutations, external effects, and delegated authority.
  \item \textbf{\underline{\emph{Runtime abstraction.}}} We introduce \stxs over result objects as a task-level validation and commit unit for tool-using agents. The abstraction groups tool intents, lineage, reversible state, staged effects, approvals, and audit metadata into transaction-scoped validation rules.
  \item \textbf{\underline{\emph{New system.}}} We build \system, a transactional execution runtime that implements this abstraction around tool dispatch. \system combines a transaction manager, shadow state, an effect outbox, and a recovery log to stage, validate, commit, abort, and audit agent effects.
\end{enumerate}

\section{Agent Tool Execution and Effect Boundaries}
\label{sec:background}

\subsection{Problem Definition}

\noindent\textbf{{Execution history.}} We model an agent as producing a long-running execution history:
\begin{equation}
  H = \langle e_1, e_2, \ldots \rangle .
\end{equation}
Each event $e_i$ is an observation, decision, tool invocation, result production, state change, approval, validation decision, or external effect produced as the agent interacts with the system and the outside world.

\noindent\textbf{{Dependency semantics.}} For events $e_i$ and $e_j$ in $H$, we write $e_i \leadsto_H e_j$ when information or state produced by $e_i$ influences the argument, payload, decision, mutation, validation outcome, or sink of $e_j$.
Let $\leadsto_H^*$ denote the transitive closure of this dependency relation.
For a side effect $s$, its dependency set is:
\begin{equation}
  \mathrm{Dep}_H(s) = \{ e_i \mid e_i \leadsto_H^* s \}.
\end{equation}
A cross-step semantic side effect is a side effect whose commit decision depends on evidence distributed across $\mathrm{Dep}_H(s)$, not on $s$ or any single prior event alone.
Secret-derived messages, untrusted-input-to-configuration writes, and rollback-sensitive multi-file edits are all examples: the relevant evidence is composed flow, not an isolated tool call.

\noindent\textbf{{Commit semantics.}} We write $\mathrm{commit}(s,t)$ when a side effect $s$ becomes durable or externally visible at time $t$, beyond the runtime's automatic rollback boundary.
A local mutation commits when it is promoted to the real workspace.
An external effect commits when it is dispatched to a service, user, network endpoint, or API.
The constraints governing commits may change over time:
\begin{equation}
  C_t = (I_t, A_t, P_t),
\end{equation}
where $I_t$ is the user's current intent, $A_t$ is the authority granted to the agent, and $P_t$ is system policy.
An execution history violates the effect boundary if it commits a side effect that does not satisfy those constraints:
\begin{equation}
  \begin{aligned}
    \exists s,t:\ 
    \mathrm{commit}(s,t) 
    & {}\land
    \neg \mathrm{Allowed}(s, C_t).
  \end{aligned}
  \label{eq:effect-boundary-violation}
\end{equation}

% \[
%   \begin{aligned}
%     H \text{ violates the boundary iff } & \exists s,t :
%     \mathrm{commit}(s,t) \\
%     & {}\land s \not\models C_t .
%   \end{aligned}
% \]

\noindent\textbf{{Boundary projections.}} An enforcement mechanism observes only a projection of the execution history.
For mechanism $M$, let $\pi_M(H_{\le t})$ denote the view of the prefix of $H$ visible to $M$ at time $t$.
Two histories are indistinguishable to $M$ when their projections are equal:
\begin{equation}
  H \equiv_M H' \iff \pi_M(H) = \pi_M(H') .
\end{equation}
If a boundary-violating history $H$ and a valid history $H'$ are indistinguishable under $\pi_M$, then $M$ cannot reliably decide whether the corresponding side effect should commit.
The classification below describes existing mechanisms by where they place the effect boundary and which projection of the agent execution they observe.

\subsection{Execution Projections and Boundaries}

Tool-using LLM agents turn model outputs into concrete system actions that read private state, run commands, edit files, call APIs, and send messages.
To understand where current mechanisms succeed and where gaps remain, we surveyed work on agent safety~\cite{debenedetti2024agentdojo,andriushchenko2024agentharm}, prompt-injection resistance~\cite{wallace2024instructionHierarchy,debenedetti2024agentdojo}, tool mediation~\cite{openaiAgentsHitl,langchainHitl,shi2025toolhijacker}, runtime containment~\cite{dockerSeccomp,aisiInspectSandboxing}, output filtering~\cite{rebedea2023nemoGuardrails,protectaiLLMGuard}, recovery~\cite{mohan1992aries,chen2015fscq}, and skill or plugin supply-chain security~\cite{snykToxicSkills,snykAgentScan,invariantMcpScan}.
Existing mechanisms provide useful projections of agent execution, but none makes the composed task flow available as a commit and recovery unit.
We describe these projections along two dimensions:

\begin{itemize}
  \item \textbf{Commit-boundary placement.}
  This dimension locates when a mechanism decides whether an agent-produced effect may become durable or externally visible.
  We organize existing mechanisms into five stages: \textbf{\textit{pre-planning}} defenses, \textbf{\textit{tool-dispatch}} gates, \textbf{\textit{execution-time}} confinement, \textbf{\textit{pre-release}} checks, and \textbf{\textit{post-effect}} recovery or audit.
  Early stages can block effects before damage, whereas later stages can inspect more realized behavior but may no longer prevent irreversible effects.

  \item \textbf{Observed execution projection.}
  This dimension identifies the portion of the execution history a mechanism observes when it evaluates agent behavior.
  We classify projections as \textbf{\textit{linguistic evidence}}, \textbf{\textit{operational evidence}}, or \textbf{\textit{semantic-effect evidence}}.
  These projections range from prompts and model outputs to operation traces.
  A transaction-level projection further includes composed result objects, dependencies, local mutations, and pending effects produced by a task.
\end{itemize}

\begin{figure}
  \centering
  \includegraphics[width=\linewidth]{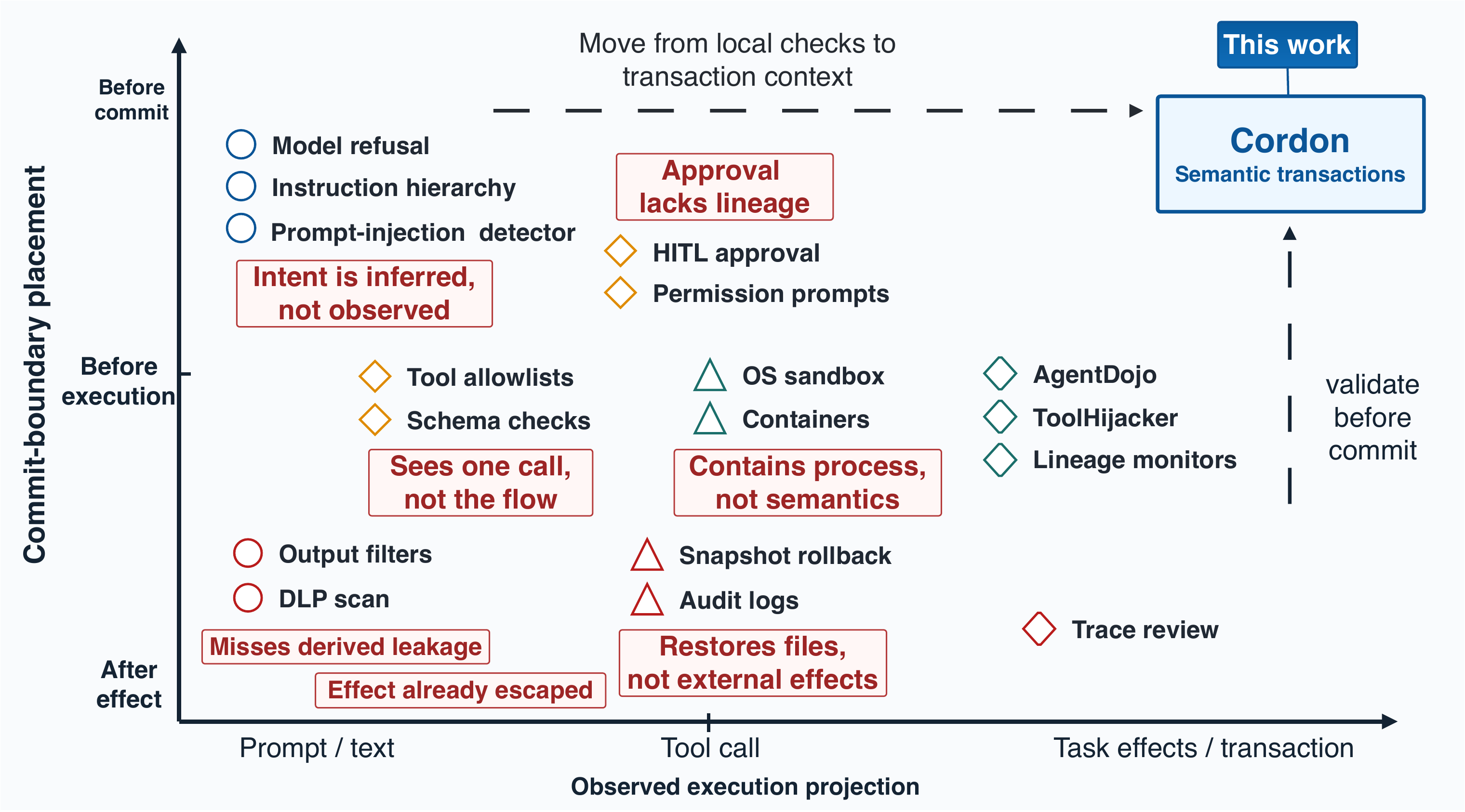}
  \caption{Existing mechanisms observe different projections of agent execution, while \system adds a task-level transaction boundary before irreversible commit.}
  \label{fig:effect-boundary-taxonomy}
\end{figure}

\subsection{Mapping Existing Boundaries}

Figure~\ref{fig:effect-boundary-taxonomy} positions representative mechanisms by the boundary they provide and the execution projection they observe before commit.

\noindent\textbf{Model and prompt defenses.}
One line of work hardens the planner against malicious or lower-priority instructions.
Alignment and instruction-hierarchy methods train models to follow intended policies and privileged instructions~\cite{ouyang2022training,bai2022constitutional,wallace2024instructionHierarchy}, while prompt-injection defenses separate or classify untrusted prompts and responses before tool use~\cite{hines2024spotlighting,chen2024struq,inan2023llamaGuard}.
These defenses act early over a planning or text projection, before concrete result objects, mutations, and effects exist.

\noindent\textbf{Tool-call gates and human approval.}
Many current agents and agent frameworks, such as the OpenAI Agents SDK~\cite{openaiAgentsHitl}, LangChain~\cite{langchainHitl}, Codex CLI~\cite{openaiCodexCli}, Claude Code~\cite{anthropicClaudeCodeSettings}, and Hermes Agent~\cite{hermesTirithSecurity}, mediate actions at the tool-call boundary.
They use allowlists, schema checks, permission prompts, and human-in-the-loop approval to decide whether a requested invocation should proceed.
This boundary is easy to integrate because tool calls are where model text becomes a system action.
However, an invocation projection rarely contains the lineage, rollback scope, pending effect set, or approval context needed to decide whether the composed task should commit.

\noindent\textbf{Runtime containment.}
Sandboxing executes agent commands and tools inside an isolated environment instead of directly in the user's local workspace.
Docker seccomp profiles and similar OS isolation mechanisms can restrict syscalls, filesystem access, and network behavior~\cite{dockerSeccomp}.
The Inspect Sandboxing Toolkit applies this idea to scalable agent evaluation~\cite{aisiInspectSandboxing}.
This process and isolation projection protects the host from high-risk execution, but it does not define which sandboxed outputs should be promoted, which effects should be released, or how derived result objects influence later commits.

\noindent\textbf{Output filters and data loss prevention.}
Data loss prevention (DLP) systems aim to detect and block sensitive information before it leaves a protected boundary.
Guardrail systems and scanners apply this idea to agents by inspecting model outputs or outbound payloads before they leave the agent boundary.
NeMo Guardrails provides programmable rails for LLM applications~\cite{rebedea2023nemoGuardrails}, and LLM Guard scans prompts and responses for risks such as prompt injection, secrets, and sensitive data~\cite{protectaiLLMGuard}.
This payload projection can catch literal leaks, but it is fragile against summarization, encoding, multi-turn context, and other derived disclosures because it lacks transaction-level lineage.

\noindent\textbf{Supply-chain scanning.}
As agents gain third-party skills, plugins, and MCP servers, installation-time trust becomes another boundary.
ToxicSkills documents malicious agent skills in the wild~\cite{snykToxicSkills}, while Snyk Agent Scan and mcp-scan inspect agent components and MCP servers before use~\cite{snykAgentScan,invariantMcpScan}.
This installation-time boundary reduces the risk of importing malicious capabilities, but it does not validate the composed behavior of a task after trusted tools begin executing.

\noindent\textbf{Recovery mechanisms.}
Storage recovery mechanisms show how systems record, restore, and reason about state after failures~\cite{mohan1992aries,chen2015fscq}.
They are central to any transactional execution substrate, but filesystem snapshots alone do not model result-object lineage or staged external effects.
For agent runtimes, this state recovery projection must be coupled to the same commit boundary that decides which mutations and effects are allowed to become visible.

\noindent\textbf{Summary.}
Taken together, these mechanisms provide useful boundaries around prompts, individual tools, processes, payloads, components, and post-effect recovery.
Our analysis highlights three remaining systems challenges.
First, decisions made before execution often lack the realized evidence needed to validate cross-step semantic side effects.
Second, decisions made after a single operation see only a fragment of the dependency chain that led to a mutation or effect.
Third, recovering from committed external effects can be costly or impossible.
These observations motivate a runtime boundary that treats tool intents, result dependencies, local mutations, staged effects, authority, and audit metadata as one transaction-level execution flow.

\section{Semantic Transaction Model}
\label{sec:model}

\subsection{Transaction Abstraction}

Existing mechanisms observe useful projections of agent execution, but do not expose the composed task flow as a commit and recovery unit.
A \emph{\stx} provides that unit.
It is a task-level transaction scope that groups model-driven tool use, semantic results, reversible state, staged effects, delegated authority, and audit evidence into a validation contract.
The model draws on transaction processing and recovery, but adapts the boundary to agent execution where effects may be semantic, cross-step, and external~\cite{haerder1983principles,mohan1992aries,garciamolina1987sagas}.

\begin{figure}[t]
  \centering
  \includegraphics[width=\columnwidth]{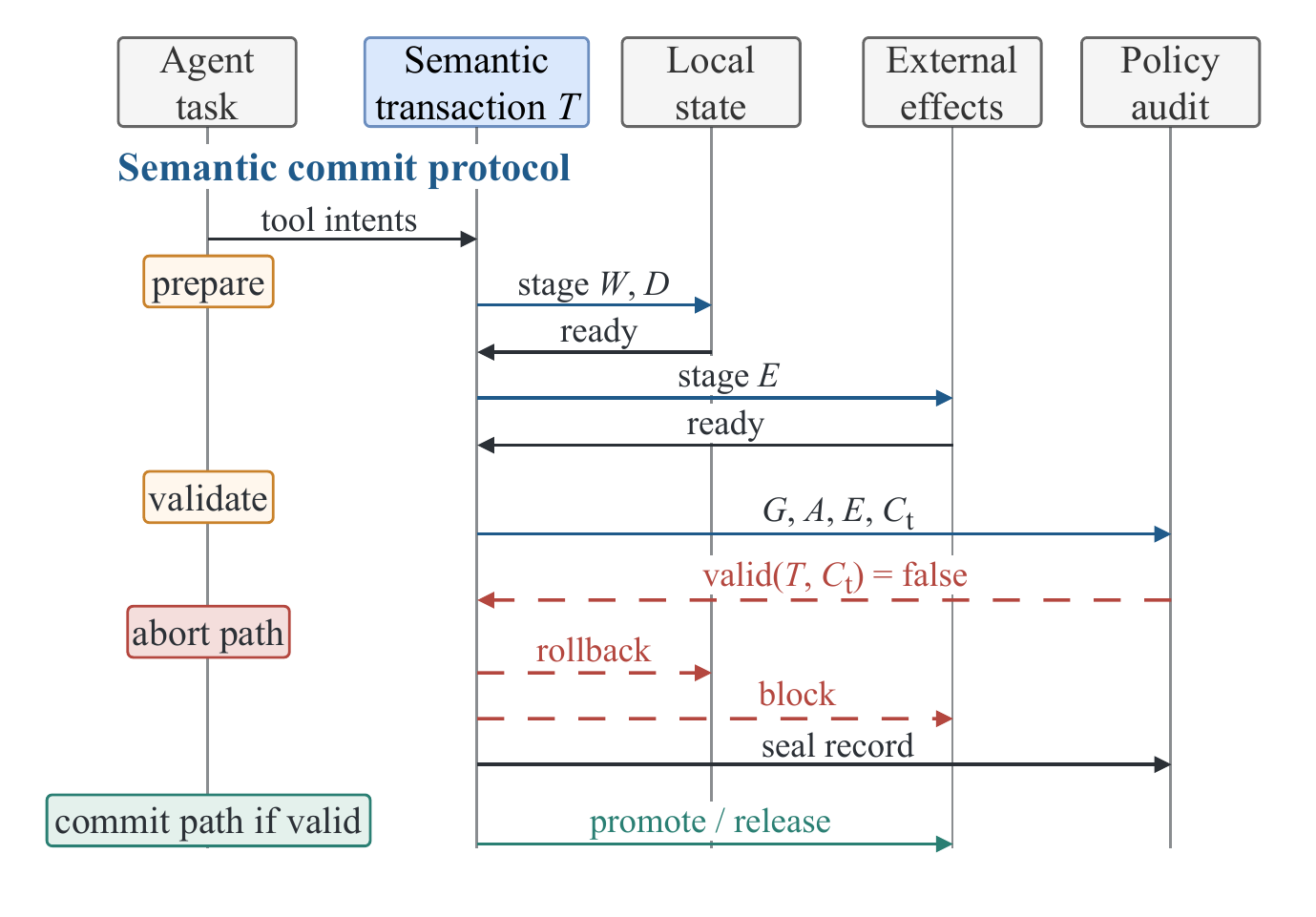}
  \caption{Semantic commit protocol for a task-level transaction boundary.}
  % \vspace{-.1in}
  \label{fig:semantic-transaction-model}
  \Description{A sequence diagram showing an agent task sending tool intents to a semantic transaction. The transaction prepares local state and external effects, validates lineage and authority with policy and audit evidence, and then either aborts or commits.}
\end{figure}

Figure~\ref{fig:semantic-transaction-model} presents the model as a semantic commit protocol over abstract domains.
The protocol has three phases:
\begin{itemize}[leftmargin=*,noitemsep,topsep=2pt]
  \item \emph{Prepare.}
  The transaction accepts tool intents, records reversible local mutations in $W \cup D$, and places external effects in $E$ without releasing them.
  \item \emph{Validate.}
  The transaction evaluates lineage $G$, authority $A$, staged effects $E$, and the constraint tuple $C_t$ as one commit unit, deriving $\mathrm{valid}(T,C_t)$.
  \item \emph{Commit/abort.}
  If $\mathrm{valid}(T,C_t)$ holds, the transaction promotes recoverable state and releases approved effects.
  Otherwise, it rolls back $W \cup D$, blocks $E$, and seals an audit record.
\end{itemize}
The lifelines define semantic domains that participate in the commit decision without prescribing an implementation architecture. Formally, a semantic transaction can be represented as:
\begin{equation}
  T = \langle scope, intents, R, W, D, E, O, G, A, status \rangle .
\end{equation}
The \textit{scope} identifies the delegated task whose effects are evaluated together.
\textit{intents} denotes the typed operations requested during that scope.
$R$ contains anchors read or observed during execution.
$O$ contains semantic result objects.
$G$ records dependencies among observations, results, mutations, and effects.
$W$ and $D$ contain recoverable writes and deletes.
$E$ contains external effects that remain staged until commit.
$A$ represents delegated authority and approval obligations.
\textit{status} records the lifecycle state of the transaction.

\begin{table}[t]
\centering
\small
\caption{Semantic transaction fields.}
% \vspace{-.1in}
\Description{A table defining the fields of a semantic transaction with colored role labels for boundary, evidence, lineage, state, effects, authority, and lifecycle.}
\label{tab:transaction-fields}
\definecolor{TxTableModelBlue}{HTML}{1F5A8A}
\definecolor{TxTableStateTeal}{HTML}{2A7F73}
\definecolor{TxTableStagedAmber}{HTML}{C9822B}
\definecolor{TxTableSlateInk}{HTML}{2B3036}
\definecolor{TxTablePaperGray}{HTML}{EEF1F3}
\definecolor{TxTableBlueWash}{HTML}{E8F1F7}
\definecolor{TxTableAmberWash}{HTML}{F7EAD6}
\newcommand{\fieldrole}[3]{%
  \begingroup
  \setlength{\fboxsep}{1.5pt}%
  \colorbox{#1}{\strut\textcolor{#2}{\scriptsize\bfseries #3}}%
  \endgroup
}
\renewcommand{\arraystretch}{1.08}
\begin{tabularx}{\columnwidth}{@{}>{\raggedright\arraybackslash}p{0.16\columnwidth}>{\centering\arraybackslash}p{0.14\columnwidth}>{\raggedright\arraybackslash}X@{}}
\toprule
Role & Field & Model meaning \\
\midrule
\fieldrole{TxTableBlueWash}{TxTableModelBlue}{Boundary} &
\textit{scope} &
Task-level transaction boundary. \\
\fieldrole{TxTableBlueWash}{TxTableModelBlue}{Boundary} &
\textit{intents} &
Typed operations within the task scope. \\
\addlinespace[1pt]
\fieldrole{TxTablePaperGray}{TxTableSlateInk}{Evidence} &
$R$ &
Anchors read or observed during execution. \\
\addlinespace[1pt]
\fieldrole{TxTableBlueWash}{TxTableModelBlue}{Lineage} &
$O$ &
Tool-returned or derived semantic objects. \\
\fieldrole{TxTableBlueWash}{TxTableModelBlue}{Lineage} &
$G$ &
Dependencies across results, state, and effects. \\
\addlinespace[1pt]
\fieldrole{TxTableBlueWash}{TxTableStateTeal}{State} &
$W,D$ &
Recoverable local writes and deletes. \\
\fieldrole{TxTableAmberWash}{TxTableStagedAmber}{Effects} &
$E$ &
External effects staged until commit. \\
\addlinespace[1pt]
\fieldrole{TxTablePaperGray}{TxTableSlateInk}{Authority} &
$A$ &
Delegated authority and approval obligations. \\
\fieldrole{TxTablePaperGray}{TxTableSlateInk}{Lifecycle} &
\textit{status} &
Active, validating, committed, aborted, compensated. \\
\bottomrule
\end{tabularx}
\end{table}

\begin{table*}[t]
\centering
\small
\caption{Transactional containment invariants.}
% \vspace{-.1in}
\label{tab:invariants}
\definecolor{InvTableModelBlue}{HTML}{1F5A8A}
\definecolor{InvTableStagedAmber}{HTML}{C9822B}
\definecolor{InvTableValidatePurple}{HTML}{6F4EB2}
\definecolor{InvTableSlateInk}{HTML}{2B3036}
\definecolor{InvTableBlueWash}{HTML}{E8F1F7}
\definecolor{InvTableAmberWash}{HTML}{F7EAD6}
\definecolor{InvTablePurpleWash}{HTML}{EFE8F7}
\newcommand{\invgroup}[3]{%
  \begingroup
  \setlength{\fboxsep}{1.5pt}%
  \colorbox{#1}{\strut\textcolor{#2}{\scriptsize\bfseries #3}}%
  \endgroup
}
\renewcommand{\arraystretch}{1.08}
\begin{tabularx}{\textwidth}{@{}>{\centering\arraybackslash}p{0.055\textwidth}>{\raggedright\arraybackslash}p{0.135\textwidth}>{\raggedright\arraybackslash}p{0.225\textwidth}>{\raggedright\arraybackslash}X@{}}
\toprule
\textbf{ID} & \textbf{Group} & \textbf{Invariant} & \textbf{Commit-condition reading} \\
\midrule
\textbf{I1} & \invgroup{InvTableBlueWash}{InvTableModelBlue}{Flow} & Secret-derived sink flow &
No path in $G$ connects secret-bearing or secret-derived $O$ to an unauthorized external sink. \\
\textbf{I7} & \invgroup{InvTableBlueWash}{InvTableModelBlue}{Flow} & Lineage preservation &
Lineage follows every transformation that preserves semantic dependence across intermediate objects. \\
\addlinespace[1pt]
\textbf{I2} & \invgroup{InvTableAmberWash}{InvTableStagedAmber}{Commit} & Untrusted sensitive mutation &
Untrusted inputs cannot justify sensitive local mutations unless $C_t$ validates the transaction. \\
\textbf{I3} & \invgroup{InvTableAmberWash}{InvTableStagedAmber}{Commit} & Pre-commit irreversible effect &
Each $e \in E$ remains staged until validation succeeds and the transaction commits. \\
\textbf{I4} & \invgroup{InvTableAmberWash}{InvTableStagedAmber}{Commit} & Rollback correctness &
Abort, denial, timeout, or rollback restores every reversible anchor in $W \cup D$. \\
\textbf{I5} & \invgroup{InvTableAmberWash}{InvTableStagedAmber}{Commit} & Boundary completeness &
Every side-effecting operation is represented in $T$ before commit or recorded as a boundary violation. \\
\addlinespace[1pt]
\textbf{I6} & \invgroup{InvTablePurpleWash}{InvTableValidatePurple}{Authority} & Least-privilege authority &
Granted authority remains within the transaction's intent, resource, sink, capability, and time scope. \\
\textbf{I8} & \invgroup{InvTablePurpleWash}{InvTableValidatePurple}{Authority} & Scoped human approval &
Approval binds to one transaction object, action, sink, and time window. \\
\textbf{I9} & \invgroup{InvTablePurpleWash}{InvTableValidatePurple}{Authority} & Audit completeness &
Commit decisions, effect transitions, approvals, violations, and recovery actions leave durable metadata. \\
\bottomrule
\end{tabularx}
\end{table*}

A transaction may span several model turns, intermediate artifacts, retries, and derived prompts when those steps serve the same delegated task.
Conversely, a new task or a change in delegated authority creates a new transaction scope.
An active transaction may move to validation, commit, abort, or compensation.
Committed external effects cannot later be treated as ordinary recoverable state.
The transaction therefore defines both the unit of local rollback and the unit of evidence for external accountability.

\subsection{Object Semantics}

The model distinguishes three object classes.
A \emph{\resultobj} is any value returned to or derived within agent execution, including file contents, tool outputs, command stdout or stderr, summaries, temporary artifacts, encoded data, and final-response candidates.
A \emph{mutation} is a local write, delete, configuration change, or persistence change over an anchor in $W \cup D$.
Mutations are recoverable only within the transaction scope that contains them.
An \emph{\effectobj} is an action whose commit makes information or behavior externally visible, such as a message, network request, issue tracker update, API call, or final output to the user.

This distinction separates state that can be rolled back from effects that must be delayed, audited, or compensated.
Local mutations have rollback semantics because the transaction can restore their pre-transaction anchors.
External effects have commit semantics because their release may cross an irreversible boundary.
The model does not require all effects to be dangerous.
It requires the commit decision to know whether an effect is still staged, whether it has authority, and whether its derivation is valid.

\subsection{Lineage Semantics}

The lineage graph $G$ records semantic dependency.
An edge $o_i \rightarrow o_j$ means that result object $o_j$ was computed from, summarized from, decoded from, or otherwise influenced by $o_i$.
An edge $o \rightarrow w$ or $o \rightarrow d$ means that result object $o$ justifies a local mutation.
An edge $o \rightarrow e$ means that result object $o$ influences the payload, destination, decision, or authorization context of effect object $e$.
Lineage is broader than string containment.
A result may be secret-derived even if it no longer contains an exact secret literal.

Lineage edges are typed by the object classes they connect.
Edges from $R$ to $O$ record observations that entered the task.
Edges within $O$ record transformations such as summarization, extraction, decoding, synthesis, or formatting.
Edges from $O$ to $W \cup D$ and $E$ record how earlier results justify local mutations or external effects.
This typed graph gives the commit rule a semantic dependency structure rather than a bag of tool logs.

\paragraph{Example mapping.}
The incident-response example maps to a compact lineage path:
$o_{\mathit{log}} \rightarrow o_{\mathit{summary}} \rightarrow e_{\mathit{slack}}$.
Here $o_{\mathit{log}} \in O$ is secret-bearing, $o_{\mathit{summary}}$ is derived, $e_{\mathit{slack}} \in E$ is pending, and $A$ determines whether the external sink is authorized.
The commit decision depends on this path, not on the Slack payload or any single tool call alone.

\subsection{Commit and Recovery Semantics}
The commit decision uses the constraint tuple:
\begin{equation}
  C_t = (I_t, A_t, P_t),
\end{equation}
where $I_t$ is the user's current intent, $A_t$ is the authority granted to the agent, and $P_t$ is policy.
A transaction $T$ is valid when it satisfies the validation contract under the current constraints:
\begin{equation}
  \mathrm{valid}(T, C_t).
\end{equation}
The model admits commit only after the transaction is valid, external effects are still staged, and reversible state has a recovery boundary:
\begin{equation}
  \mathrm{commit}(T) \Rightarrow
  \mathrm{valid}(T,C_t) \land \mathrm{staged}(E) \land \mathrm{recoverable}(W,D).
\end{equation}

Commit promotes recoverable local state and releases approved staged effects.
Abort discards recoverable local mutations and prevents staged effects from becoming visible.
Rollback returns $W \cup D$ to the pre-transaction scope.
Recovery preserves enough durable evidence to distinguish committed, aborted, pending, and compensated effects after a failure.
The model gives ACID-like rollback semantics only for recoverable local state.
External effects in $E$ are different: they are delayed until validation and commit, assigned idempotency and audit metadata for release, and treated as audit or compensation cases once their external boundary has been crossed~\cite{garciamolina1987sagas}.

\subsection{Transactional Containment Invariants}

The invariants in Table~\ref{tab:invariants} define the commit-validity contract for a semantic transaction.
They specify what must hold before transaction state or effects may commit.
The invariants fall into three groups: lineage flow, commit discipline, and authority/accountability.

The invariant suite is not a complete theory of agent safety.
It defines transaction-level conditions for deciding whether agent-produced state and effects may commit.
I1 and I7 capture semantic result flow.
I2 through I5 capture commit discipline for reversible and irreversible effects.
I6 through I9 bind privileges, approvals, and accountability to the transaction object.
These invariants also define violation oracles for workloads that test transaction-level containment.

The model-to-system mapping is direct but separate from the semantics.
In \system, the transaction object maps to a runtime transaction context; staged effects map to an effect outbox; recoverable writes map to shadow state; lineage maps to durable metadata; and authority obligations map to approval and validation machinery.
This mapping is a realization choice, not part of the semantic validity definition.

\section{Runtime Architecture}
\label{sec:design}

\begin{figure*}
  \centering
  \includegraphics[width=\textwidth]{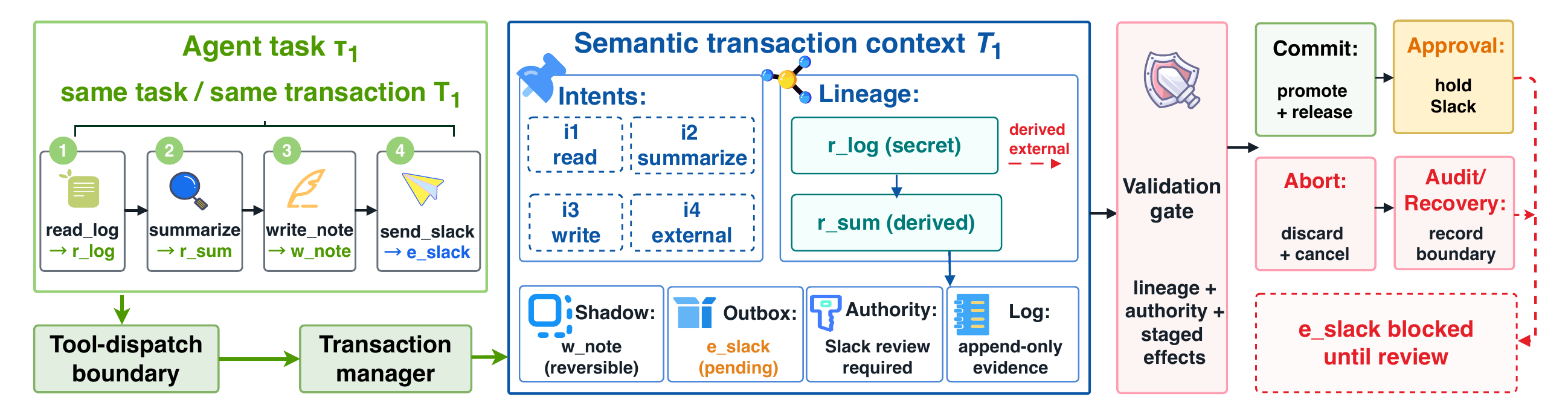}
  % \vspace{-.2in}
  \caption{\system runtime architecture.}
  \label{fig:architecture}
\end{figure*}

\system realizes the semantic transaction model by inserting a transactional control plane between the agent planner and side-effecting tools.
The architecture is operational: it binds tool calls to task transactions, materializes runtime evidence, executes local changes speculatively, stages external effects, validates the composed transaction, and records enough state for crash recovery and audit.

\subsection{Runtime Overview}

Figure~\ref{fig:architecture} shows the runtime path for one incident-response task with four tool calls.
The agent reads logs, summarizes the result, writes a remediation note, and prepares a Slack message.
\system places a mediation layer at the tool-dispatch boundary, where tool arguments are concrete but side effects have not yet committed.
A transaction manager creates or resumes the active transaction for the task and attaches each mediated operation to that context.

The center of the figure is the runtime transaction context.
It is the concrete carrier for the model state defined in Section~\ref{sec:model}: operation intents, object handles, lineage edges, speculative local state, pending external effects, scoped authority, and recovery records.
These entries are maintained as the task runs rather than reconstructed from the final prompt or final output.
The validation engine consumes this accumulated context and produces a transaction outcome.
Commit promotes approved local state and releases approved effects.
Abort discards speculative local state and cancels pending effects.
Approval holds a scoped action without broadening authority.
Audit and recovery records preserve crossed-boundary evidence.

\begin{table}[t]
\centering
\footnotesize
\caption{\system runtime mechanisms and the model properties they preserve.}
\Description{A table mapping Cordon runtime mechanisms to design challenges and semantic model properties.}
\label{tab:design-mechanisms}
\renewcommand{\arraystretch}{1.05}
\begin{tabularx}{\columnwidth}{@{}>{\raggedright\arraybackslash}p{0.24\columnwidth}>{\raggedright\arraybackslash}p{0.36\columnwidth}>{\raggedright\arraybackslash}X@{}}
\toprule
Design challenge & Runtime mechanism & Model property preserved \\
\midrule
\textbf{Task binding} &
Mediation layer and transaction manager bind calls to a task context. &
Transaction scope and complete intent set. \\
\textbf{Evidence materialization} &
Object handles and metadata records are appended online. &
Lineage graph and validation evidence. \\
\textbf{Speculative state} &
Shadow-state engine executes local writes outside the real workspace. &
Recoverable $W,D$ and no uncommitted visibility. \\
\textbf{External effects} &
Effect outbox holds sink, payload, lineage, and release status. &
Staged $E$, idempotence, and scoped approval. \\
\textbf{Failure handling} &
Recovery log records commits, aborts, releases, and boundary crossings. &
Recoverability, auditability, and compensation state. \\
\bottomrule
\end{tabularx}
\end{table}

\subsection{Tool-Dispatch Interposition}

\system interposes where the agent runtime dispatches tool calls.
This point has concrete tool names, arguments, resources, and destinations, but it precedes real filesystem mutations, network sends, and service updates.
The mediation layer parses each call into a runtime operation and asks the transaction manager for the active task context.
Side-effecting operations must carry a transaction identifier before they execute.
Unsupported tools, opaque plugins, and calls that cannot describe their resources or effects are rejected or recorded as boundary violations.

Interposition also stabilizes task identity across the messy execution patterns of agents.
An agent may issue calls over several turns, retry failed operations, or interleave reads with speculative writes.
The transaction manager stores the active transaction binding with the task context and requires later calls to reuse that binding.
This prevents an operation that consumes earlier task results from appearing as an unrelated action at validation time.
Read-only operations can execute with low overhead, but their returned handles enter the transaction context once later operations derive state or effects from them.

\subsection{Online Transaction Construction}

\system constructs transaction evidence incrementally while the task executes.
When a tool returns, the runtime assigns a handle to the result and records the handle in the transaction context.
When a later operation summarizes that result, passes it to a command, writes it into local state, or uses it to build an external payload, the metadata layer appends the corresponding dependency edge.
The validation engine therefore sees the concrete execution history that produced the pending state and effects.

This online construction avoids treating the model's memory as the source of truth.
Command stdout, file contents, summaries, artifacts, and candidate final responses can all become tracked objects even when the model later paraphrases them.
The context stores handles and edges, not model rationales.
As a result, validation can inspect the task's operational history without replaying the model's reasoning or trusting a final payload to reveal every dependency.

\vspace{-.1in}
\subsection{Shadow-State Engine}

Local mutations execute through a shadow-state engine.
File writes, deletes, command-produced changes, and supported configuration updates are applied to a transaction-scoped view rather than the real workspace.
Subsequent reads consult this view, so the agent can observe its speculative work while uncommitted changes remain invisible outside the transaction.
At commit, the engine promotes approved writes and deletes to the real workspace.
At abort, denial, timeout, or rollback, it discards the shadow view and restores the pre-transaction anchors.

The shadow-state boundary gives \system ACID-like behavior only for local state that the engine can mediate.
Tools that mutate state through unsupported channels cannot receive ordinary rollback semantics.
For those operations, the mediation layer either blocks execution or records the unsupported boundary as recovery evidence.
This choice keeps the runtime honest: an operation is either inside the recoverable state boundary or explicitly treated as a crossed boundary.

\subsection{Effect Outbox}

External actions enter an effect outbox instead of dispatching immediately.
Each outbox entry records the sink, payload handle, lineage handle, authority state, idempotency key, and release status.
The validation engine evaluates these entries together with the transaction context.
If validation succeeds and the required authority is present, commit releases approved entries and records their dispatch state.
If validation fails, abort cancels entries that remain pending.
If user approval is required, the outbox holds only the scoped action under review.

The commit protocol separates local promotion from external release.
Local state can be promoted atomically within the mediated workspace.
External effects require dispatch tracking because the runtime cannot physically undo the external world after release.
The outbox therefore records whether an effect is pending, released, aborted, or manually recovered.
Idempotency keys prevent duplicate dispatch during crash recovery, while release records let audit distinguish a blocked effect from an effect that crossed the boundary.

\subsection{Commit and Recovery Protocol}

\system treats commit as a two-part protocol over mediated state and staged effects.
Before commit, the runtime writes a commit manifest that names the staged local changes, scheduled deletes, pending effects, validation token, and target promotion scope.
The manifest is the recovery point for the interval between validation success and durable transaction completion.
Local promotion and effect release are then resolved separately.
Local state is promoted only if the manifest still matches the staged state, while external effects remain in the outbox until the commit path marks them ready for release.

Crash recovery is driven by the transaction state recorded before the crash.
A prepared transaction with no durable execution receipts can abort automatically.
An executing transaction with partial receipts requires manual review because the runtime cannot prove how far command execution progressed.
A validating transaction returns to validation because no commit decision has become durable.
A precommit transaction is reconciled from its manifest: if staged files and targets are still consistent, \system completes promotion and finalizes the transaction; if they are inconsistent, \system rolls back from the recorded recovery boundary.
Committed, aborted, and rolled-back transactions are treated as terminal except for unfinished outbox effects.

The protocol is conservative for effects that may have crossed an external boundary.
Pending or ready effects remain pending after recovery.
Cancelled effects remain cancelled.
Dispatched effects are not resent unless the runtime has idempotency evidence that the previous dispatch did not complete.
Otherwise, the transaction reaches an audit or compensation state that records the sink, payload lineage, authority state, and observed dispatch status.
This distinction lets \system provide automatic recovery for mediated local state without pretending that irreversible external behavior can always be physically undone.

% \vspace{-.1in}
\section{Implementation}

We implement \system as a research prototype with roughly 14.4 KLOC of handwritten runtime code and a 3.9 KLOC benchmark driver.
Most of the runtime is written in Python, including the adapter, validation service, operation service, and shared transaction contracts.
The cross-process interface is specified in a small protobuf/gRPC IDL, which generates Python client and server stubs.
For command containment, the prototype links against a Rust-backed nono sandbox through its Python bindings.

\noindent\textbf{RPC interface.}
We use gRPC~\cite{grpcDocs} to separate framework-facing adapters from the local service that owns durable state and side-effect mediation.
The protobuf file exposes two services.
The validation-facing service provides compile, validate, result-evaluation, lineage-query, and status RPCs.
The operation-facing service provides prepare, execute, commit, approval, effect-acknowledgement, rollback, recovery, query, and status RPCs.
All RPCs carry a JSON envelope that contains serialized transaction contract objects.
This choice keeps the interface stable while the transaction contract evolves and localizes field-level validation in the Python contract layer.

\noindent\textbf{State management.}
We use ZODB FileStorage~\cite{zodbDocs} as the transactional metadata store.
This lets the prototype persist Python objects directly while still using explicit commit and abort operations around each state transition.
Large artifacts are kept out of the object database.
Snapshots, sandbox directories, shadow files, and commit manifests live as files under the \system data root, while ZODB stores object identifiers, paths, content digests, status fields, and recovery records.
The idempotency table stores one response per RPC operation and idempotency key.
This makes retries safe for prepare, execute, commit, rollback, approval resolution, effect acknowledgement, and recovery after client or daemon restarts.

\noindent\textbf{Shadow filesystem.}
We implement shadow state as an application-level manifest system.
Each committed turn has a directory of staged file contents and a JSON manifest mapping absolute workspace paths to staged files or delete tombstones.
Task and session views are rebuilt by replaying ordered turn manifests into aggregate manifests.
For command execution, we materialize a temporary runtime directory by copying the real working tree and then overlaying task and session shadow entries.
This userspace design runs across macOS and Linux.
Command startup is proportional to the materialized workspace size.
To keep the cost manageable, sandbox snapshots exclude common high-churn directories such as \textit{.git}, \textit{node\_modules}, \textit{\_\_pycache\_\_}, and Python bytecode files.
We canonicalize all paths before anchoring them and acquire resource locks in sorted anchor order to avoid deadlock during concurrent promotion.

\noindent\textbf{Sandbox backend.}
We build command containment on top of a nono-style sandbox backend~\cite{nonoSandbox}.
The wrapper constructs a capability set for each command.
It grants read access to required system paths, grants read-write access only to the materialized runtime directory, and blocks network access by default.
When a tool declaration requires network access, the wrapper starts a local proxy with an allowed-host list and credential routes.
Proxy audit events are converted into egress records.
After execution, the sandbox snapshot manager reports created, modified, permission-changed, and deleted paths.
The wrapper maps those runtime paths back to the observed workspace, hashes stdout and stderr, stores an execution trace, and converts changed files into staged writes.

\noindent\textbf{Framework integration.}
The framework-facing code is intentionally small.
The adapter layer contains an SDK, an async RPC client, a task runtime, a server manager, an interceptor, and one host-runtime bridge.
The bridge leaves the model loop unchanged.
It replaces the normal tool invocation function with a wrapper that records the tool name, arguments, task identifier, working directory, and authority context before forwarding the call through \system.
Porting \system to another Python agent framework requires implementing the same wrapper plus a capability-spec mapping from framework tools to declared reads, writes, effects, and approval requirements.
The middleware uses tool-call metadata and capability specifications as its integration contract.

\noindent\textbf{Engineering tradeoffs.}
Several implementation choices favor portability and observability over raw performance.
The JSON-envelope RPC interface keeps schema evolution in the Python contract layer.
The manifest-based shadow filesystem runs entirely in userspace and materializes command views by copying files.
The object database gives crash-consistent metadata updates for the prototype.
High-throughput distributed deployment would require a storage backend designed for remote coordination.
The sandbox wrapper inherits the platform coverage of the underlying nono backend and reports backend availability during startup.
These tradeoffs match the current goal: a concrete prototype that exposes the transaction boundary precisely enough to evaluate containment, rollback, audit, and integration cost.

\section{Evaluation}
\label{sec:evaluation}

We evaluate \system as a transaction boundary for agent tool side effects.
The evaluation measures whether \system contains cross-step semantic risks, preserves practical usability, maintains practical system performance, and provides correct transaction behavior under rollback and crash recovery.

\subsection{Experimental Setup}

\noindent\textbf{Benchmark suites.}
We use two benchmark suites with the same runtime configuration, tool interfaces, policy configuration, and transaction logging format.
The shared correlated-risk suite contains 45 risk-bearing multi-tool workflows constructed as the cross product of nine defense-boundary categories and five transaction-level risk families. 
The workloads span six agent domains---coding, incident response, document processing, office workflows, customer support, and data-analysis tasks---and cover sensitive writes, exec-mediated sensitive writes, session-secret external effects, derived-secret exec egress, and high-fanout deletes. 
The workflows are constructed from task patterns and operational structures commonly used in existing agent benchmarks and evaluations for software engineering and knowledge-work agents~\cite{yang2024sweagent,drouin2024workarena,debenedetti2024agentdojo,andriushchenko2024agentharm,shi2025toolhijacker}.
% The task domains follow prior agent evaluations in software engineering and knowledge-work settings~\cite{yang2024sweagent,drouin2024workarena}.
% The risk construction follows agent prompt-injection and harmful-agent benchmarks~\cite{debenedetti2024agentdojo,andriushchenko2024agentharm,shi2025toolhijacker}, but labels transaction-level side effects rather than final-answer harm alone.
We use this suite both for containment and for performance, where the same workflows run under plain execution and three transaction-mediated approval policies: approve, reject, and mixed.
The rollback suite contains five deterministic failed-agent trajectories that execute a bad agent step, roll it back, and run a resume check.
We additionally use $\tau$-bench~\cite{yao2024taubench} and Terminal-Bench~\cite{merrill2026terminal} as benign task-completion sanity checks.

\noindent\textbf{Metrics.}
For containment, we report pre-commit interception and policy-violating effects.
For performance, we report task and total wall time, token usage, LLM calls, and approval events.
For rollback, we report rollback latency, total recovery time, residual deltas, and resume success.
For standard benign benchmarks, we report the official task success or correctness score.

\noindent\textbf{Baselines.}
We compare \system against six baselines, each corresponding to a missing transaction capability and a common control point in current agent systems:
\begin{itemize}[leftmargin=*]
  \item \textbf{Plain Agent:} no semantic transaction mediation, commit boundary, or recovery protocol.
  \item \textbf{Per-Tool Policy:} static rules over tool names and arguments, without result-object lineage or composed-task validation.
  \item \textbf{Human Approval:} tool-level approval, without scoped transaction approvals over objects, sinks, and time windows~\cite{openaiAgentsHitl,langchainHitl}.
  \item \textbf{Sandbox Only:} command/process containment, without commit semantics, staged effects, or lineage-aware promotion~\cite{aisiInspectSandboxing,nonoSandbox}.
  \item \textbf{Snapshot Rollback:} filesystem or Git snapshot recovery, without staged external effects or transaction-level effect acknowledgments~\cite{haerder1983principles,mohan1992aries,garciamolina1987sagas}.
  \item \textbf{Output Filter:} regex, secret scanner, or DLP-style outbound filtering, without lineage across transformations and turns~\cite{inan2023llamaGuard,rebedea2023nemoGuardrails,protectaiLLMGuard}.
\end{itemize}

We run each baseline on the suites where it has meaningful semantics.
For example, output filtering applies to containment and usability but has no rollback protocol, while snapshot rollback applies to local recovery but cannot retract a dispatched external effect.

\noindent\textbf{Platform and measurement.}
We integrate \system into a commercial tool-using agent runtime, which we refer to as \emph{Agent-H}.
All large-model calls in the experiments use DeepSeek-V4-Pro~\cite{deepseekai2026deepseekv4}.
Unless stated otherwise, decoding parameters, tool set, and \system policy remain fixed across systems.
Each run records a transaction trace with tool calls, result labels, lineage edges, staged effects, approvals, recovery events, and terminal state.

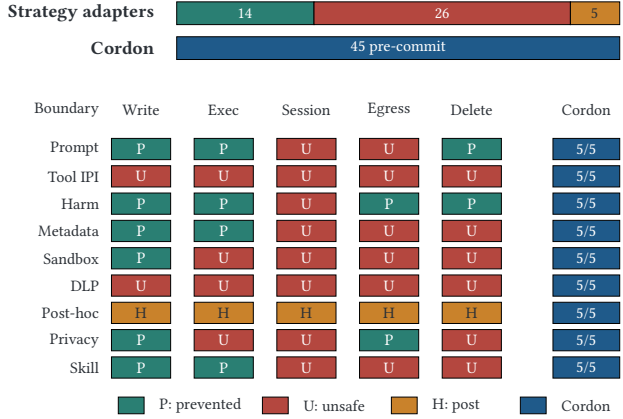
\begin{figure}[t]
  \centering
  \resizebox{\columnwidth}{!}{%
    \definecolor{CordonBlue}{HTML}{1F5A8A}
\definecolor{RuntimeTeal}{HTML}{2A7F73}
\definecolor{StagedAmber}{HTML}{C9822B}
\definecolor{PolicyRed}{HTML}{B4473F}
\definecolor{SlateInk}{HTML}{2B3036}
\definecolor{PaperGray}{HTML}{EEF1F3}
\definecolor{BlueWash}{HTML}{E8F1F7}
\definecolor{AmberWash}{HTML}{F7EAD6}
\definecolor{RedWash}{HTML}{F4DEDC}

\begin{tikzpicture}[
  x=1cm,
  y=1cm,
  every node/.style={font=\scriptsize, text=SlateInk},
  tinylabel/.style={font=\tiny, text=SlateInk},
  rowlabel/.style={font=\tiny, text=SlateInk, anchor=east},
  toplabel/.style={font=\scriptsize\bfseries, text=SlateInk, anchor=east},
  axis/.style={draw=SlateInk!55, line width=0.35pt},
  cell/.style={minimum width=0.78cm, minimum height=0.28cm, inner sep=0pt, draw=black, line width=0.28pt},
  prevented/.style={cell, fill=RuntimeTeal},
  missed/.style={cell, fill=PolicyRed},
  posthoc/.style={cell, fill=StagedAmber},
  cordon/.style={cell, fill=CordonBlue},
  barprev/.style={fill=RuntimeTeal, draw=black, line width=0.28pt},
  barunsafe/.style={fill=PolicyRed, draw=black, line width=0.28pt},
  barpost/.style={fill=StagedAmber, draw=black, line width=0.28pt},
  barcordon/.style={fill=CordonBlue, draw=black, line width=0.28pt}
]

\path[use as bounding box] (0.34,-0.10) rectangle (7.90,5.55);

% Overall stacked bars.
\node[toplabel] at (1.88,5.28) {Strategy adapters};
\node[toplabel] at (1.88,4.82) {\system};
\draw[barprev] (2.05,5.12) rectangle ++(1.81,0.30);
\draw[barunsafe] (3.86,5.12) rectangle ++(3.38,0.30);
\draw[barpost] (7.24,5.12) rectangle ++(0.65,0.30);
\node[tinylabel, text=white] at (2.95,5.27) {14};
\node[tinylabel, text=white] at (5.55,5.27) {26};
\node[tinylabel, text=SlateInk] at (7.56,5.27) {5};
\draw[barcordon] (2.05,4.66) rectangle ++(5.84,0.30);
\node[tinylabel, text=white] at (4.97,4.81) {45 pre-commit};

% Matrix headers.
\node[tinylabel, anchor=east] at (1.15,4.00) {Boundary};
\node[tinylabel, align=center] at (1.58,4.00) {Write};
\node[tinylabel, align=center] at (2.67,4.00) {Exec};
\node[tinylabel, align=center] at (3.76,4.00) {Session};
\node[tinylabel, align=center] at (4.85,4.00) {Egress};
\node[tinylabel, align=center] at (5.94,4.00) {Delete};
\node[tinylabel, align=center] at (7.45,4.00) {\system};

% Row labels.
\foreach \y/\lab in {
  3.48/Prompt,
  3.12/Tool IPI,
  2.76/Harm,
  2.40/Metadata,
  2.04/Sandbox,
  1.68/DLP,
  1.32/Post-hoc,
  0.96/Privacy,
  0.60/Skill
} {
  \node[rowlabel] at (1.15,\y) {\lab};
}

% Matrix cells. P = prevented before commit, U = unsafe committed, H = post-hoc only.
\newcommand{\pcell}[2]{\node[prevented] at (#1,#2) {\textcolor{white}{\tiny P}};}
\newcommand{\ucell}[2]{\node[missed] at (#1,#2) {\textcolor{white}{\tiny U}};}
\newcommand{\hcell}[2]{\node[posthoc] at (#1,#2) {\textcolor{SlateInk}{\tiny H}};}
\newcommand{\ccell}[2]{\node[cordon, minimum width=0.88cm] at (#1,#2) {\textcolor{white}{\tiny 5/5}};}

% Prompt input: P P U U P.
\pcell{1.58}{3.48} \pcell{2.67}{3.48} \ucell{3.76}{3.48} \ucell{4.85}{3.48} \pcell{5.94}{3.48} \ccell{7.45}{3.48}
% Tool IPI: U U U U U.
\ucell{1.58}{3.12} \ucell{2.67}{3.12} \ucell{3.76}{3.12} \ucell{4.85}{3.12} \ucell{5.94}{3.12} \ccell{7.45}{3.12}
% Harm guard: P P U P P.
\pcell{1.58}{2.76} \pcell{2.67}{2.76} \ucell{3.76}{2.76} \pcell{4.85}{2.76} \pcell{5.94}{2.76} \ccell{7.45}{2.76}
% Tool metadata: P P U U U.
\pcell{1.58}{2.40} \pcell{2.67}{2.40} \ucell{3.76}{2.40} \ucell{4.85}{2.40} \ucell{5.94}{2.40} \ccell{7.45}{2.40}
% Sandbox: P U U U U.
\pcell{1.58}{2.04} \ucell{2.67}{2.04} \ucell{3.76}{2.04} \ucell{4.85}{2.04} \ucell{5.94}{2.04} \ccell{7.45}{2.04}
% Output/DLP: U U U U U.
\ucell{1.58}{1.68} \ucell{2.67}{1.68} \ucell{3.76}{1.68} \ucell{4.85}{1.68} \ucell{5.94}{1.68} \ccell{7.45}{1.68}
% Post-hoc trace: H H H H H.
\hcell{1.58}{1.32} \hcell{2.67}{1.32} \hcell{3.76}{1.32} \hcell{4.85}{1.32} \hcell{5.94}{1.32} \ccell{7.45}{1.32}
% Privacy lineage: P U U P U.
\pcell{1.58}{0.96} \ucell{2.67}{0.96} \ucell{3.76}{0.96} \pcell{4.85}{0.96} \ucell{5.94}{0.96} \ccell{7.45}{0.96}
% Skill policy: P P U U U.
\pcell{1.58}{0.60} \pcell{2.67}{0.60} \ucell{3.76}{0.60} \ucell{4.85}{0.60} \ucell{5.94}{0.60} \ccell{7.45}{0.60}

% Legend uses the same wash-fill/accent-outline treatment as the matrix.
\node[prevented, minimum width=0.34cm, minimum height=0.22cm] at (1.45,0.10) {};
\node[tinylabel, anchor=west] at (1.70,0.10) {P: prevented};
\node[missed, minimum width=0.34cm, minimum height=0.22cm] at (3.35,0.10) {};
\node[tinylabel, anchor=west] at (3.60,0.10) {U: unsafe};
\node[posthoc, minimum width=0.34cm, minimum height=0.22cm] at (5.05,0.10) {};
\node[tinylabel, anchor=west] at (5.30,0.10) {H: post};
\node[cordon, minimum width=0.34cm, minimum height=0.22cm] at (6.75,0.10) {};
\node[tinylabel, anchor=west] at (7.00,0.10) {\system};

\end{tikzpicture}%
  }
  \caption{Security detection results across all correlated-risk cases.}
  \label{fig:additional-containment-matrix}
\end{figure}

\subsection{Security and Containment}

% \noindent\textbf{Containment setup.}
% We use the shared correlated-risk suite described above to compare pre-commit containment across defense boundaries.

\noindent\textbf{Case-level containment result.}
Plain execution commits the risky effect in all 45 cases.
Strategy adapters derived from existing defense boundaries prevent 14 cases before commit, miss 26 cases, and detect 5 cases only after commit.
\system intercepts all 45 cases before commit.
Figure~\ref{fig:additional-containment-matrix} shows both the aggregate result and the case-level structure.
Each matrix cell is one benchmark case.
Rows correspond to the defense boundary exposed to the strategy adapter; columns correspond to the transaction-level risk exercised by the case.

\noindent\textbf{Failure modes.}
The matrix shows three recurring failure modes.
Local-view defenses, such as prompt-level guards, tool metadata checks, and harmful-action guards, prevent some direct or visibly dangerous writes, but they become unstable when the write is framed as routine workflow metadata or when the sensitive mutation occurs through an allowed command.
Effect-boundary defenses, such as sandbox policies and output guardrails, protect useful local boundaries, yet they do not see the full relation between an approved secret read and a later message, command upload, or workspace mutation.
Post-hoc trace monitors identify the risky execution after it happens, but the irreversible effect has already crossed the boundary.

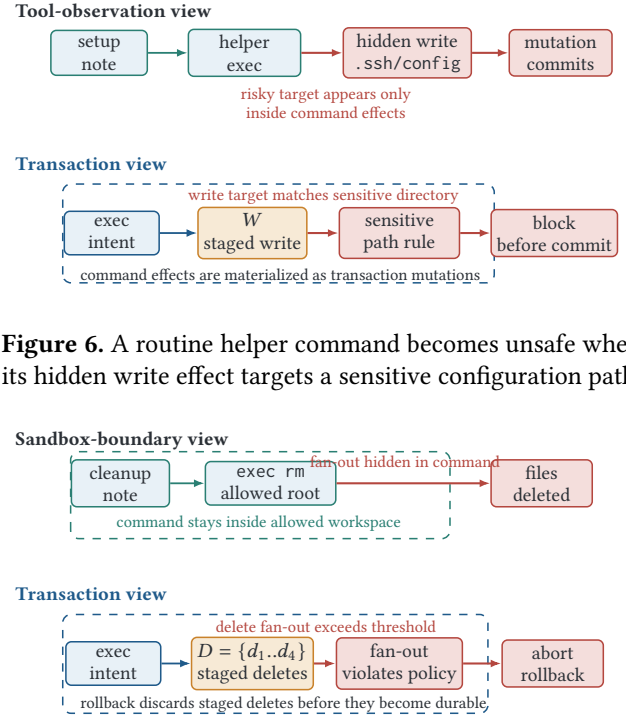
\begin{figure}[t]
  \centering
  \definecolor{CordonBlue}{HTML}{1F5A8A}
\definecolor{RuntimeTeal}{HTML}{2A7F73}
\definecolor{StagedAmber}{HTML}{C9822B}
\definecolor{PolicyRed}{HTML}{B4473F}
\definecolor{SlateInk}{HTML}{2B3036}
\definecolor{BlueWash}{HTML}{E8F1F7}
\definecolor{AmberWash}{HTML}{F7EAD6}
\definecolor{RedWash}{HTML}{F4DEDC}

\resizebox{\columnwidth}{!}{%
\begin{tikzpicture}[
  x=1cm,
  y=1cm,
  every node/.style={font=\scriptsize, text=SlateInk},
  title/.style={font=\scriptsize\bfseries, text=SlateInk, anchor=west},
  note/.style={font=\tiny, align=center, text=SlateInk},
  box/.style={rounded corners=2pt, line width=0.55pt, align=center,
    minimum height=0.55cm, inner sep=2pt},
  local/.style={box, draw=RuntimeTeal, fill=BlueWash, minimum width=1.22cm},
  object/.style={box, draw=CordonBlue, fill=BlueWash, minimum width=1.20cm},
  staged/.style={box, draw=StagedAmber, fill=AmberWash, minimum width=1.30cm},
  bad/.style={box, draw=PolicyRed, fill=RedWash, minimum width=1.30cm},
  dep/.style={-{Latex[length=1.35mm]}, draw=CordonBlue, line width=0.65pt},
  risk/.style={-{Latex[length=1.35mm]}, draw=PolicyRed, line width=0.75pt},
  safe/.style={-{Latex[length=1.35mm]}, draw=RuntimeTeal, line width=0.65pt},
  boundary/.style={draw=CordonBlue, dashed, rounded corners=3pt, line width=0.55pt}
]

\path[use as bounding box] (-0.05,-2.00) rectangle (8.05,1.95);

\node[title] at (0.00,1.72) {Tool-observation view};
\node[local] (note1) at (1.18,1.18) {setup\\note};
\node[local, minimum width=1.42cm] (exec1) at (3.02,1.18) {helper\\exec};
\node[bad, minimum width=1.62cm] (write1) at (5.08,1.18) {hidden write\\\texttt{.ssh/config}};
\node[bad] (commit1) at (7.02,1.18) {mutation\\commits};
\draw[safe] (note1) -- (exec1);
\draw[risk] (exec1) -- (write1);
\draw[risk] (write1) -- (commit1);
\node[note, text=PolicyRed] at (4.05,0.55) {risky target appears only\\inside command effects};

\node[title, text=CordonBlue] at (0.00,-0.22) {Transaction view};
\draw[boundary] (0.72,-1.73) rectangle (6.18,-0.48);
\node[object] (intent) at (1.34,-1.10) {exec\\intent};
\node[staged] (delta) at (3.12,-1.10) {$W$\\staged write};
\node[bad, minimum width=1.52cm] (rule) at (4.98,-1.10) {sensitive\\path rule};
\node[bad] (block) at (6.94,-1.10) {block\\before commit};
\draw[dep] (intent) -- (delta);
\draw[risk] (delta) -- (rule);
\draw[risk] (rule) -- (block);
\node[note, text=PolicyRed] at (4.02,-0.63) {write target matches sensitive directory};
\node[note, anchor=west] at (0.82,-1.62) {command effects are materialized as transaction mutations};

\end{tikzpicture}%
}
  \vspace{-.2in}
  \caption{A routine helper command becomes unsafe when its hidden write effect targets a sensitive configuration path.}
  \label{fig:case-study-exec-sensitive-write}
  % \Description{A two-lane diagram for an exec-mediated sensitive-write case. The upper lane shows a tool-observation view that sees a setup note and helper exec while the command writes to an SSH config path. The lower lane shows Cordon materializing the command effect as a staged write and blocking it before commit because it targets a sensitive directory.}
\end{figure}

\noindent\textbf{Case studies.}
Figure~\ref{fig:case-study-exec-sensitive-write} shows an exec-mediated sensitive-write case.
The agent reads a setup note and runs a routine-looking helper command while triaging a checkout issue.
The unsafe behavior is not visible in the abstract tool request: the command appends data to an SSH configuration path.
A tool-observation boundary sees a setup note and a helper command, while \system materializes the command's write effect as a staged transaction mutation and blocks it before commit.

Figure~\ref{fig:case-study-highfanout-delete} shows a high-fanout deletion case.
The cleanup command stays inside an allowed workspace, so a sandbox-style boundary can treat the command as locally permitted.
The transaction view exposes the operation as a staged delete set whose fan-out exceeds the policy threshold.
\system aborts the transaction and discards the staged deletes before they become durable.

\begin{figure}[t]
  \centering
  \definecolor{CordonBlue}{HTML}{1F5A8A}
\definecolor{RuntimeTeal}{HTML}{2A7F73}
\definecolor{StagedAmber}{HTML}{C9822B}
\definecolor{PolicyRed}{HTML}{B4473F}
\definecolor{SlateInk}{HTML}{2B3036}
\definecolor{PaperGray}{HTML}{EEF1F3}
\definecolor{BlueWash}{HTML}{E8F1F7}
\definecolor{AmberWash}{HTML}{F7EAD6}
\definecolor{RedWash}{HTML}{F4DEDC}

\resizebox{\columnwidth}{!}{%
\begin{tikzpicture}[
  x=1cm,
  y=1cm,
  every node/.style={font=\scriptsize, text=SlateInk},
  title/.style={font=\scriptsize\bfseries, text=SlateInk, anchor=west},
  note/.style={font=\tiny, align=center, text=SlateInk},
  box/.style={rounded corners=2pt, line width=0.55pt, align=center,
    minimum height=0.55cm, inner sep=2pt},
  local/.style={box, draw=RuntimeTeal, fill=BlueWash, minimum width=1.24cm},
  staged/.style={box, draw=StagedAmber, fill=AmberWash, minimum width=1.24cm},
  bad/.style={box, draw=PolicyRed, fill=RedWash, minimum width=1.28cm},
  object/.style={box, draw=CordonBlue, fill=BlueWash, minimum width=1.20cm},
  dep/.style={-{Latex[length=1.35mm]}, draw=CordonBlue, line width=0.65pt},
  risk/.style={-{Latex[length=1.35mm]}, draw=PolicyRed, line width=0.75pt},
  safe/.style={-{Latex[length=1.35mm]}, draw=RuntimeTeal, line width=0.65pt},
  sandbox/.style={draw=RuntimeTeal, dashed, rounded corners=3pt, line width=0.55pt},
  boundary/.style={draw=CordonBlue, dashed, rounded corners=3pt, line width=0.55pt}
]

\path[use as bounding box] (-0.05,-2.00) rectangle (8.05,1.95);

\node[title] at (0.00,1.72) {Sandbox-boundary view};
\draw[sandbox] (0.82,0.48) rectangle (5.62,1.58);
\node[local] (note1) at (1.45,1.18) {cleanup\\note};
\node[local, minimum width=1.65cm] (exec1) at (3.35,1.18) {\texttt{exec rm}\\allowed root};
\node[bad] (del1) at (6.78,1.18) {files\\deleted};
\draw[safe] (note1) -- (exec1);
\draw[risk] (exec1) -- (del1);
\node[note, text=PolicyRed] at (5.05,1.46) {fan-out hidden in command};
\node[note, text=RuntimeTeal] at (3.20,0.68) {command stays inside allowed workspace};

\node[title, text=CordonBlue] at (0.00,-0.22) {Transaction view};
\draw[boundary] (0.72,-1.73) rectangle (6.10,-0.48);
\node[object] (intent) at (1.36,-1.10) {exec\\intent};
\node[staged, minimum width=1.42cm] (dset) at (3.12,-1.10) {$D=\{d_1..d_4\}$\\staged deletes};
\node[bad, minimum width=1.48cm] (valid) at (4.98,-1.10) {fan-out\\violates policy};
\node[bad] (abort) at (6.92,-1.10) {abort\\rollback};
\draw[dep] (intent) -- (dset);
\draw[risk] (dset) -- (valid);
\node[note, text=PolicyRed] at (4.05,-0.63) {delete fan-out exceeds threshold};
\draw[risk] (valid) -- (abort);
\node[note, anchor=west] at (0.82,-1.62) {rollback discards staged deletes before they become durable};

\end{tikzpicture}%
}
  \vspace{-.2in}
  \caption{A workspace-local command can still be unsafe when its composed transaction effect is a high-fanout delete.}
  \label{fig:case-study-highfanout-delete}
  \Description{A two-lane diagram for a high-fanout delete case. The upper lane shows a sandbox-boundary view that allows an exec command inside the workspace. The lower lane shows a semantic transaction that records a staged delete set, detects that the fan-out exceeds policy, and aborts before commit.}
\end{figure}

\noindent\textbf{Why semantic transactions intercept.}
\system does not rely on a stronger single-point classifier at the prompt, tool, or output boundary.
It records result lineage, stages local and external effects, and validates the composed execution flow before commit.
This lets \system block cases where the decisive evidence is distributed across earlier context, derived artifacts, command side effects, and pending external actions.

\subsection{System Performance}

\begin{figure*}[t]
  \centering
  \includegraphics[width=\textwidth]{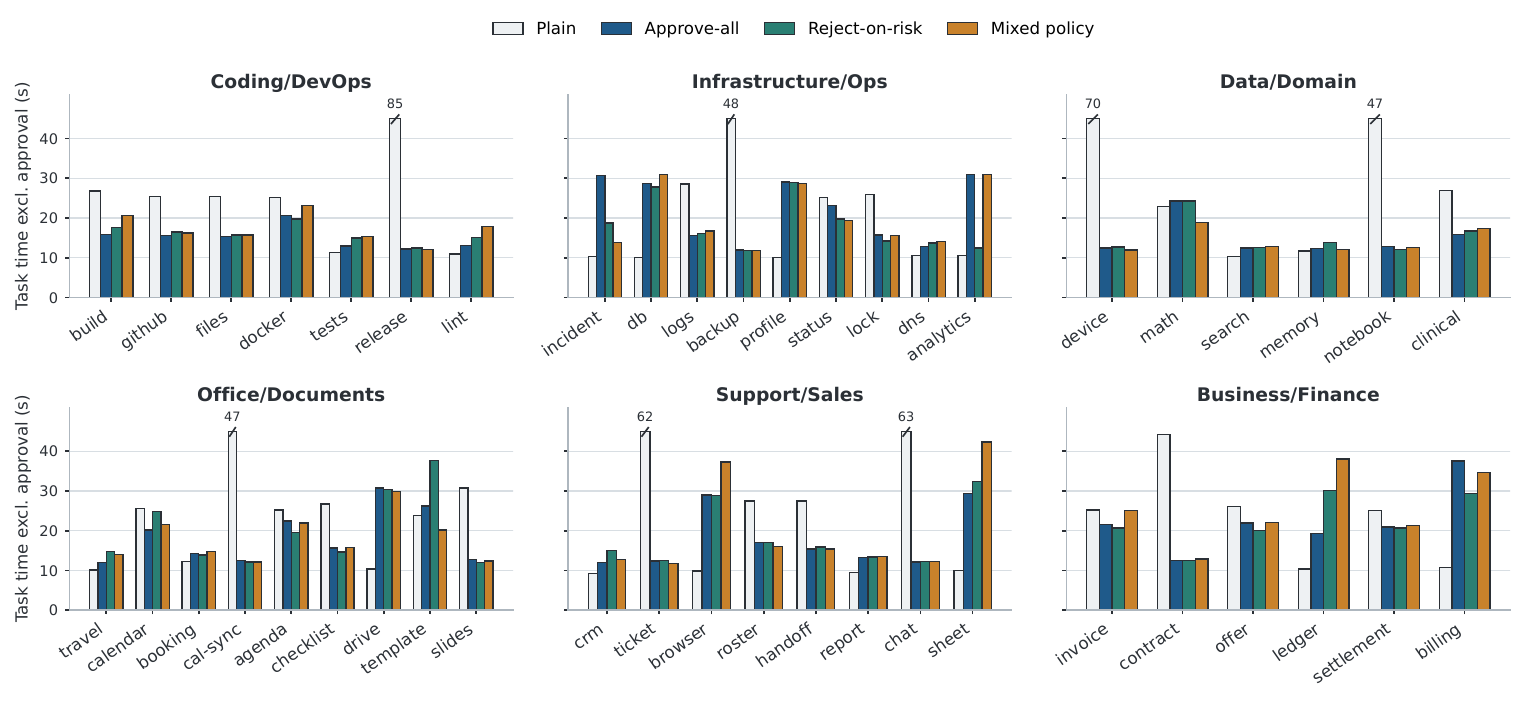}
  \vspace{-.2in}
  \caption{End-to-end task time including approval wait for all 45 workflows under plain execution and three transaction-mediated approval policies.}
  \label{fig:system-overhead-bars}
  \Description{A two-row, three-column grouped bar chart. Each panel corresponds to an application workflow class and contains grouped bars for individual workflows. Each workflow has four bars: plain, approve-all, reject-on-risk, and mixed policy. Bar heights include approval wait time. The y-axis is truncated at 45 seconds, and capped bars show numeric labels for their measured values.}
\end{figure*}

\noindent\textbf{End-to-end performance.} Table~\ref{tab:agent-overhead} summarizes aggregate end-to-end behavior across all 45 workflows, while Figure~\ref{fig:system-overhead-bars} reports the per-workflow distribution.
The measurement includes approval wait so that the end-to-end numbers reflect the current prototype path.

\begin{table}[t]
\centering
\small
\caption{End-to-end performance on 45 risk-bearing multi-tool agent workflows. Task times include approval wait; Appr. reports approval events.}
\Description{A table comparing plain execution with three transaction-mediated approval modes on mean task time, median task time, LLM calls, token usage, and approval events.}
\label{tab:agent-overhead}
\renewcommand{\arraystretch}{1.08}
\begin{tabular*}{\columnwidth}{@{\extracolsep{\fill}}lrrrrr@{}}
\toprule
Mode & Mean & Med. & Calls & Tokens & Appr. \\
\midrule
Plain & 25.55s & 25.16s & 162 & 1.89M & 0 \\
Approve & 31.35s & 30.74s & 119 & 1.36M & 45 \\
Reject & 23.64s & 21.88s & 125 & 1.42M & 36 \\
Mixed & 31.12s & 28.74s & 127 & 1.45M & 40 \\
\bottomrule
\end{tabular*}
\end{table}

Table~\ref{tab:agent-overhead} shows the aggregate cost of transaction mediation when approval wait is counted.
Reject-on-risk improves mean task time from 25.55s to 23.64s, a 7.5\% reduction, because rejected transactions terminate risky flows before the agent completes later tool steps.
Approve-all and mixed add approval cost, but the increase remains moderate: approve-all raises mean task time by 5.80s, or 22.7\%, and mixed raises it by 5.57s, or 21.8\%.
At the same time, all transaction-mediated modes reduce model work: token use drops by 23.6--28.4\%, and LLM calls fall from 162 to 119--127.

Figure~\ref{fig:system-overhead-bars} explains why these averages do not imply a uniform speedup or slowdown.
The fastest transaction-mediated cases are the derived-secret egress workflows, where validation stops the run before a long unsafe execution path.
In these cases, plain execution often continues through command construction and follow-up reasoning, while \system reaches a transaction boundary after fewer tool and model steps.
This effect appears across multiple application panels rather than in one isolated domain.
The slower cases are mostly rollback-heavy high-fanout delete workflows, where \system records staged effects, aborts the transaction, and lets the agent complete recovery-oriented follow-up steps.
The figure therefore shows that the overhead comes from concrete transaction work and approval handling, while the savings come from cutting off long unsafe executions before they accumulate more model calls.

\noindent\textbf{Rollback performance.} Table~\ref{tab:rollback-overhead} isolates rollback cost from model variability by using deterministic trajectories with no LLM calls.
Rollback median measures only the recovery primitive.
Recovery median measures the failed-step, rollback, and resume-check path, excluding workspace setup.
Residual deltas report the median number of filesystem differences that remain after rollback.
Resume passed counts trials whose continuation check succeeds.

\begin{table}[t]
\centering
\small
\caption{Rollback performance on deterministic failed-agent trajectories.}
% \vspace{-.1in}
% \Description{A table comparing speculative transaction rollback with Git-style recovery baselines on rollback median latency, recovery median time, residual deltas, and resume success.}
\label{tab:rollback-overhead}
\renewcommand{\arraystretch}{1.08}
\begin{tabularx}{\columnwidth}{@{}l>{\centering\arraybackslash}X>{\centering\arraybackslash}X>{\centering\arraybackslash}X>{\centering\arraybackslash}X@{}}
\toprule
\textbf{Mode} & \textbf{Rollback median} & \textbf{Recovery median} & \textbf{Residual deltas} & \textbf{Resume passed} \\
\midrule
\system & 4.17ms & 178.95ms & 0 & 15/15 \\
\texttt{restore} & 10.99ms & 100.12ms & 73 & 0/15 \\
\texttt{reset} & 12.27ms & 100.38ms & 73 & 0/15 \\
\texttt{reset+clean} & 21.74ms & 111.47ms & 0 & 12/15 \\
\bottomrule
\end{tabularx}
\end{table}

Speculative transaction rollback remains in the millisecond range, with a 4.17ms median rollback latency across 15 trials.
The Git restore and reset baselines appear inexpensive, but they leave a median of 73 residual deltas because untracked artifacts and staged effect traces remain outside the tracked-file boundary.
Reset plus clean removes those artifacts in most tasks, but it still fails the permission-drift trajectory because ordinary Git state does not restore the required file-mode state.
The transaction rollback path pays for manifest and snapshot bookkeeping, and it binds that work to the transaction scope that produced the staged mutations.

\begin{figure}[t]
  \centering
  \includegraphics[width=\columnwidth]{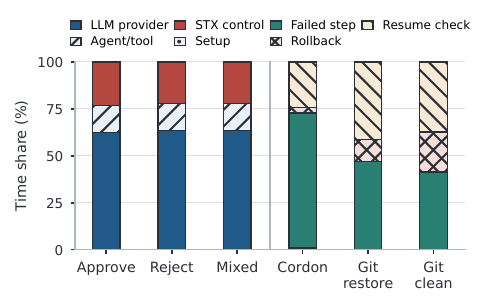}
  % \vspace{-.2in}
  \caption{Runtime cost composition for end-to-end transaction-mediated workflows and deterministic recovery paths. Bars exclude approval wait and show normalized component shares, not absolute latency.}
  \label{fig:system-overhead-breakdown}
\end{figure}

\noindent\textbf{Cost breakdown.} Figure~\ref{fig:system-overhead-breakdown} decomposes measured runtime into normalized component shares.
For end-to-end workflows, excluding approval wait, provider latency dominates the measured path: 62.3\%, 63.5\%, and 63.6\% for approve, reject, and mixed runs.
Agent and tool execution account for 14.2--14.4\%, while Cordon's transaction-control path accounts for 22.2--23.4\%.
For deterministic recovery, \system spends only 2.9\% of measured time in the rollback primitive; most time is spent executing the failed step and checking that the resumed task can continue.
The Git restore and Git clean bars spend a larger share in rollback itself, at 11.8\% and 21.4\%, because recovery is performed by applying workspace-level restore or cleanup operations during the measured rollback path.
\system shifts most of that work into transaction preparation and manifest tracking before failure, so rollback can discard staged state by transaction scope instead of reconstructing the workspace from a broad file-level operation.

\subsection{Task Correctness on Standard Benchmarks}

We use $\tau$-bench~\cite{yao2024taubench} and Terminal-Bench~\cite{merrill2026terminal} as benign task-completion checks for ordinary tool-agent behavior. 
% Specifically, we select 20 representative tasks from $\tau$-bench and 10 representative tasks from Terminal-Bench, and repeat each evaluation twice. 
Both experiments use the benchmark's official correctness metric and compare the plain agent with \system under the same model, prompt, tool interface, decoding parameters, and environment.
% \system only stages effects, logs lineage, validates transactions, and commits the same tool calls; it does not add attack-specific policies or artificial refusal rules.

\begin{table}[t]
\centering
\small
\caption{Task correctness on benign benchmark subsets.}
\Description{A table comparing the plain agent and Cordon on standard benign agent benchmarks using official task success or correctness scores.}
\label{tab:standard-correctness}
\renewcommand{\arraystretch}{0.9}
\begin{tabular*}{\columnwidth}{@{\extracolsep{\fill}}lccc@{}}
\toprule
Benchmark & Plain Agent & \system & $\Delta$ \\
\midrule
$\tau$-bench & 87.5\% & 90.0\% & +2.5 \\
Terminal-Bench & 100.0\% & 100.0\% & +0.0 \\
\bottomrule
\end{tabular*}
\end{table}

Table~\ref{tab:standard-correctness} shows that \system remains within measurement variance of the plain agent on both benchmarks while preserving the same externally evaluated task outcomes. 
Across standard benign workloads, transaction mediation, staged execution, and validation do not measurably degrade benchmark-visible correctness or prevent the agent from completing ordinary multi-step tasks. 
These results suggest that the runtime overhead introduced by semantic transactions primarily affects execution structure and containment behavior rather than the agent’s ability to solve general-purpose tasks.

% \vspace{-.2in}
% \subsection{Limitation and Future Work}

% \system provides transactional containment only for operations that remain inside its mediated execution boundary. Effects and state that flow through the transaction runtime can be staged, validated, rolled back, and audited before commit.

% The model does not extend to arbitrary opaque behavior outside the mediated boundary. 
% External services, dynamically changing tools, unsupported plugins, or operations whose resources and side effects cannot be described through capability specifications may escape semantic tracking or recovery. 
% In these cases, \system treats the operation as a crossed boundary and records the associated lineage, authority state, and recovery evidence for audit or compensation rather than assuming full containment or reversibility.

% Finally, the current prototype prioritizes explicit containment semantics over aggressive optimization. 
% Approval handling remains a major source of latency and interaction cost, especially for workflows with repeated external effects or fine-grained authority checks. 
% Many optimizations remain possible, including scoped approval reuse, approval batching, policy precomputation, incremental validation, and tighter integration with host runtimes and tool frameworks. 
% These improvements would reduce runtime overhead while preserving the transaction boundary exposed by semantic transactions.

% \subsection{Limitations and Scope}
\subsection{Limitations and Future Work}

\system does not attempt to provide complete semantic correctness or universal containment for arbitrary agent behavior. Instead, it introduces a practical runtime transaction boundary for staging and validating multi-step agent effects before irreversible commit. Its guarantees therefore apply only to operations that remain within the mediated transaction runtime and whose mutations and effects are observable to the system.

Operations that cross external or opaque boundaries, such as unsupported plugins, dynamically changing services, or tools with unobservable side effects, may fall outside this containment scope. In these cases, \system records lineage, authority context, and recovery metadata for audit and compensation rather than assuming full reversibility or semantic visibility.

Future work includes extending mediation coverage across heterogeneous tool ecosystems and reducing runtime overhead through techniques such as scoped approval reuse, incremental validation, and tighter integration with host runtimes and tool frameworks.

\section{Related Work}

\textbf{Transactions and recovery.}
Database and storage systems use logging, commit protocols, and recovery rules to decide when speculative state becomes durable~\cite{haerder1983principles,mohan1992aries,chen2015fscq}.
Coordination avoidance ties coordination to application invariants~\cite{bailis2014coordination}, while sagas handle long-running effects through compensation~\cite{garciamolina1987sagas}.
\system applies these ideas to a different transaction object: agent tool intents, result lineage, shadow state, staged effects, authority, and audit metadata.

\textbf{Provenance and flow.}
Data provenance explains how derived artifacts depend on earlier inputs and transformations~\cite{buneman2001why,cheney2009provenance}.
Information-flow systems constrain sensitive data movement across program boundaries~\cite{goguen1982security,enck2010taintdroid,zeldovich2006secureservices}.
\system does not prove full noninterference; it uses lineage as runtime commit evidence for task-level state and effects.

\textbf{Agent runtimes and benchmarks.}
ReAct and Toolformer showed that language models can interleave reasoning with tool calls~\cite{yao2023react,schick2023toolformer}.
Subsequent systems and benchmarks study agents in software engineering, workplace, multi-turn API, and cross-application settings~\cite{yang2024sweagent,drouin2024workarena,trivedi2024appworld,yao2024taubench,merrill2026terminal}.
These works make tool use a first-class capability and evaluation target, while \system targets the runtime substrate that stages, validates, commits, recovers, and audits tool-call consequences.

\textbf{Agent safety.}
Prompt-injection defenses, instruction hierarchies, harmful-agent benchmarks, human approval, sandboxing, output guardrails, tool-attack studies, and supply-chain scanners identify important risks in tool-using agents~\cite{wallace2024instructionHierarchy,hines2024spotlighting,chen2024struq,debenedetti2024agentdojo,andriushchenko2024agentharm,shi2025toolhijacker,openaiAgentsHitl,langchainHitl,dockerSeccomp,rebedea2023nemoGuardrails,protectaiLLMGuard,snykToxicSkills,snykAgentScan}.
Section~\ref{sec:background} analyzes their execution projections; \system instead provides a commit and recovery abstraction across prompts, tools, local state, external effects, and delegated authority.

\textbf{Reference monitors.}
Reference-monitor designs require mediation to be complete, tamper resistant, and small enough to reason about~\cite{anderson1972computer,lampson1974protection}.
Capability systems restrict transformations over protected objects, and Clark--Wilson emphasizes well-formed transactions~\cite{clark1987comparison}.
\system inherits this mediation principle, but its unit is a task-level semantic transaction rather than a syscall, permission check, or tool invocation.

% \vspace{-.3in}
\section{Conclusion}

We present \system, a transactional execution runtime for tool-using LLM agents. 
\system introduces semantic transactions as a task-level execution boundary that groups tool intents, result lineage, staged local state, pending external effects, delegated authority, and recovery metadata into a single commit and rollback unit.
By interposing at runtime,
\system makes agent-produced consequences explicit before they become durable or externally visible. 
Our evaluation shows that semantic transactions improve containment and recovery for cross-step semantic risks while preserving practical usability and benchmark-visible task correctness.

\bibliographystyle{ACM-Reference-Format}
\bibliography{references}

\end{document}